# Dimensionality Engineering of Magnetic Anisotropy from Anomalous Hall Effect in Synthetic SrRuO$_3$ Crystals

*Seung Gyo Jeong[1,#], Seong Won Cho[2,#], Sehwan Song[3], Jin Young Oh[1], Do Gyeom Jeong[5], Gyeongtak Han[4], Hu Young Jeong[6], Ahmed Yousef Mohamed[7], Woo-suk Noh[8], Sungkyun Park[3], Jong Seok Lee[5], Suyoun Lee[2,*], Young-Min Kim[4,*], Deok-Yong Cho[7,*] and Woo Seok Choi[1,*]*

[1]Department of Physics, Sungkyunkwan University, Suwon 16419, Korea

[2]Center for Neuromorphic Engineering, Korea Institute of Science and Technology, Seoul 02792, Korea

[3]Department of Physics, Pusan National University, Busan 46241, Korea

[4]Department of Energy Sciences, Sungkyunkwan University, Suwon 16419, Korea

[5]Department of Physics and Photon Science, Gwangju Institute of Science and Technology (GIST), Gwangju 61005, Korea

[6]Graduate School of Semiconductor Materials and Devices Engineering, Ulsan National Institute of Science and Technology, Ulsan 44919, Korea

[7]Department of Physics, Jeonbuk National University, Jeonju 54896, Korea

[8]cCPM, Max Planck POSTECH/Korea Research Initiative, Pohang 37673, Korea

# S.G.J. and S.W.C. contributed equally to this paper.

*Email: slee_eels@kist.re.kr, youngmk@skku.edu, zax@jbnu.ac.kr, choiws@skku.edu

KEYWORDS

Low-dimensional magnetism, SrRuO$_3$, magnetic anisotropy engineering, oxide superlattice, atomic-scale epitaxy, correlated magnetic order.

ABSTRACT

Magnetic anisotropy in atomically thin correlated heterostructures is essential for exploring quantum magnetic phases for next-generation spintronics. Whereas previous studies have mostly focused on van der Waals systems, here, we investigate the impact of dimensionality of epitaxially-grown correlated oxides down to the monolayer limit on structural, magnetic, and orbital anisotropies. By designing oxide superlattices with a correlated ferromagnetic $SrRuO_3$ and nonmagnetic $SrTiO_3$ layers, we observed modulated ferromagnetic behavior with the change of the $SrRuO_3$ thickness. Especially, for three-unit-cell-thick layers, we observe a significant 1,500% improvement of coercive field in the anomalous Hall effect, which cannot be solely attributed to the dimensional crossover in ferromagnetism. The atomic-scale heterostructures further reveal the systematic modulation of anisotropy for the lattice structure and orbital hybridization, explaining the enhanced magnetic anisotropy. Our findings provide valuable insights into engineering the anisotropic hybridization of synthetic magnetic crystals, offering a tunable spin order for various applications.

Magnetic anisotropy (MA) in correlated heterostructures with varied dimensions is critical for achieving interesting quantum magnetic phases that are viable for spintronic applications. When a thin ferromagnetic layer is sandwiched between nonmagnetic layers, the effective MA ($K_{eff}$) can be characterized as $K_{eff} = K_v + K_s/t + K_d$, where $K_v$, $K_s$, and $K_d$ are the volume magnetocrystalline, interfacial magnetocrystalline, and demagnetization (shape) anisotropy, respectively, and $t$ is the thickness of the magnetic layer.[1] Whereas $K_d$ originates from the magnetic dipole-dipole interaction associated with the macroscopic geometry of the sample, $K_v$ and $K_s$ are induced by finite spin-orbit coupling with orbital hybridization. The competition among the terms in $K_{eff}$ determines the direction of the magnetic easy axis and magnitude of the coercive field ($H_c$), the control of which is essential for spintronic memory applications.[1, 2] More recently, the discovery and development of low-dimensional spintronics have required a better understanding and utilization of spin-orbit-coupling-induced MA ($K_v$ and $K_s$), especially across the dimensional crossover from 3D to 2D.[3] Studies on layered materials have shown that a sufficiently strong MA overcomes thermal fluctuations and stabilizes the long-range magnetic order in 2D.[4] However, most studies on low-dimensional MA have been conducted on materials with weak van der Waals interlayer interactions, which are unsuitable for understanding correlated materials with strong orbital hybridization. The dimensionally modulated strong orbital hybridization would provide a rich playground with multiple tuning knobs for the MA control using the designed epitaxial heterostructure.[5-7] Covalent/ionic materials offer a controllable isotropic network of orbital hybridization, enabling the design and realization of an effective dimensional crossover for MA evolution.

Synthetic oxide crystals with atomic-scale precision control offer a promising pathway for exploring the dimensionality engineering of MA and the resultant magnetic ground states. The deliberate design of oxide heterostructures enables the customization of the synthetic atomic network, dimensionality, electronic structure, and magnetic properties.[7-13] The strong interplay between the charge, spin, lattice, and orbital degrees of freedom further fosters correlated functionalities that are strongly modified by varying the spatial dimension. By utilizing perovskite $SrRuO_3$ (SRO), a strongly correlated ferromagnet, we can access the dimensional crossover of a synthetic magnetic crystal by controlling the atomic thickness of the SRO layer. To maximize its dimensionality effects, we employ a sufficiently thick $SrTiO_3$ (STO) layer as a nonmagnetic insulating spacer within the synthetic crystal, which confines the thin ferromagnetic layer. A coherent perovskite structure with strong ionic bonds ensures a suitable high-quality platform for investigating the dimensional crossover to the monolayer limit.[14-16] MA plays an essential role in the emergence of spin-related functionalities in SRO, such as current-induced magnetization switching, anomalous and topological Hall effects, and topological magnon excitation.[7, 17-19] The sizeable spin-orbit coupling of SRO introduces a finite $K_{eff}$ and the resultant perpendicular MA in the bulk.[20] Several SRO heterostructures have been shown to exhibit limited MA modulation by controlling the epitaxial strain, thickness, buffer layer, nano pattern, and stoichiometry.[21-29] However, most studies have mostly concentrated on controlling scales ranging from several nanometers to tens of nanometers, and individual experimental evidence, which poses challenges to achieving an integrated understanding. Therefore, observation of complete MA modulation and its correlation with structural perturbation, spin order, and orbital hybridization across the dimensional crossover down to atomic-scale thickness is highly necessary.

We realized dimensional crossover from 2D to 3D for the synthetic SRO crystals by varying the number of SRO-layer atomic u.c. ($x$) from 1 to 6 (Figure 1(a) and see Methods). We used $y$ = 6 or 8 u.c.-thick STO layers within the superlattice to prevent possible interlayer coupling between the SRO layers.[30] Note that STO-layer thicknesses of 6 or 8 u.c. were employed to tune the probing depth of the superlattices for spectroscopy measurements, and the variation in the STO-layer thickness in this range did not change the out-of-plane ferromagnetic properties of the superlattices, as evidenced by numerous experiments (Figure S1). Figure 1(a) shows X-ray diffraction (XRD) $\theta$-$2\theta$ scans and atomic-force microscopy representing high-quality periodic supercell structures with atomically well-defined interfaces and surfaces of the SRO/STO superlattices. As $x$ increases, the angular spacing between the superlattice peaks ($SL^{\pm n}$) decreases, accurately reproducing the expected diffraction pattern from the schematically shown structure in the right panel of Figure 1(a). The inset of Figure 1(a) demonstrates that the surface step-terrace structure is well maintained after the superlattice growth. The XRD reciprocal space map (RSM) in Figure 1(b) around the (103) plane of STO shows a fully strained state for the superlattice with the thickest SRO layer ($x$ = 6), representing the in-plane lattice constant; hence, the epitaxial in-plane strain is identical for all superlattices.

Magnetic ($H$)-field-dependent Hall resistivity ($\rho_{xy}$) at 15 K exhibits anomalous $\rho_{xy}$, revealing the dimensional crossover effect, as shown in Figure 1(c). Owing to the close correlation between the ferromagnetic spin order and anomalous $\rho_{xy}$, we can characterize the ferromagnetic contribution of the atomically thin SRO layers. We note that the $\rho_{xy}$ signals of atomically thin

SRO layers (especially for SRO layer thickness < 3 u.c.) have not been reported previously in conventional SRO single films due to their highly insulating nature. Taking advantage of the superlattice structure, a slightly elevated temperature (15 K), and the ac lock-in technique, we were able to observe $\rho_{xy}$ down to the monolayer SRO. To extract the anomalous Hall signal of the SRO layers, we subtracted the contribution of the ordinary Hall effect using linear fitting above $H_c$ (see Methods and Figure S2). The $\rho_{xy}$ ($H$) curve of the $x = 6$ superlattice shows the typical ferromagnetic hysteresis behavior with saturation $\rho_{xy}$ at 4 T ($\rho_s$) and remanent $\rho_{xy}$ at 0 T ($\rho_r$) values remaining, i.e., $\rho_r/\rho_s = 1$. Such robust (square-like) ferromagnetic behavior ($\rho_r/\rho_s = 1$) is maintained down to an $x = 3$ superlattice with the $\rho_s$ and $\rho_r$ values decreasing proportionally.[18] However, as $x$ further decreases below $x = 3$, the sign of $\rho_s$ for a given $H$-field reverses, and the relative $\rho_r$-to-$\rho_s$ values decrease significantly. The sign reversal of $\rho_s$ with decreasing film thickness has been frequently observed in SRO thin films (albeit below 4-6 u.c. for the single films), which was attributed to the Berry curvature effect.[31] The sign of the anomalous Hall signal in SRO is determined by various sources of Berry curvature near the Fermi level,[32] which can induce the SRO thickness dependent sign change of anomalous Hall effect. Figure 1(d) (black symbols, left $y$-axis) summarizes that the $\rho_r/\rho_s$ ratios deviate from 1 below $x = 3$ and eventually reach 0 for the monolayer SRO superlattice. The absence of the ferromagnetic order in monolayer SRO leads to the vanishing of $\rho_r$ at observed temperature; however, the application of a finite $H$-field may induce the angular orbital momentum of conduction electrons in 2D systems resulting in a nonzero $\rho_s$, similar to the field-induced anomalous Hall signal suggested in nonferromagnetic low-dimensional systems.[33] We note that nonferromagnetic behavior of monolayer SRO from anomalous Hall signal at 15 K differs from previous experimental reports.[16] This discrepancy may be attributed to the limited measurement temperature of the Hall effect of the insulating SRO monolayer and/or differences in the thickness of the STO layer in the SRO/STO superlattice, which sensitively determines the magnetic ground state of the monolayer SRO superlattice.[30] Temperature-dependent magnetization curves (the insets of Figure 1(d) and Figure S3(a)) show the $x$-dependent enhancement of the ferromagnetic transition temperature ($T_c$) (light blue symbols, right $y$-axis of Figure 1(d)), revealing similar $x$-dependence with the $\rho_r/\rho_s$ ratio evolution. These results consistently indicate the emergence and evolution of a robust ferromagnetic order in the SRO layers owing to the dimensional crossover from 2D to 3D.

Interestingly, $x$-dependent $H_c$ shows the highest value at $x = 3$ and it decreases with increasing or decreasing $x$, whereas the ferromagnetic behavior ($\rho_r/\rho_s$ ratio and $T_c$) enhances and saturates at $x = 3$ with increasing $x$. $H_c$ values observed in $\rho_{xy}$ ($H$) curves reflect the preference of the spin-orbit-coupling-induced uniaxial MA of ferromagnetic SRO layers within the superlattices,[19] which have a unique advantage to observe magnetic anisotropy of SRO superlattice near the monolayer limits. The $H_c$ value of the $x = 1$ superlattice is 0 owing to the absence of ferromagnetic order; however, the value significantly increases with increasing $x$ and reaches 1.59 T for the $x = 3$ superlattice at 15 K (black symbols, left $y$-axis of Figure 1(e)). In contrast, when $x$ further increases above 3, the $H_c$ value decreases systematically, down to 0.75 T for the $x = 6$ superlattice. A similar systematic decrease of the $H_c$ values above $x = 3$ was observed at 1.8 K, the temperature of which was too low to measure the $\rho_{xy}$ ($H$) of insulating $x = 1$ and 2 superlattices (Figure S4). The $H_c$ value was as high as 3.11 T for the $x = 3$ superlattice (red symbols, right $y$-axis in Figure 1(e)) at 1.8 K, which is more than 1,500% enhancement with respect to the SRO single film (~0.2 T, we plot the data as $x = 100$) at low temperature.[34, 35]

Figure S3(b) shows that the $H$-field-dependent magnetization along the out-of-plane direction consistently exhibits the same $x$-dependent $H_c$, confirming the results of $\rho_{xy}$ ($H$) measurements. The large and nonmonotonic change of $H_c$ with atomic-scale precision control implies that the dimensional crossover of ferromagnetism alone cannot provide a sufficient explanation of MA modulation. Because SRO is a system with delicate structural variations such as octahedral distortion and coupled orbital hybridization, the possible evolution of structural anisotropy near the low-dimensional limit within the SRO/STO superlattice as a function of $x$ strongly influences the MA.

To examine the local structural anisotropy at the atomic-scale two-dimensional limit, we artificially constructed SRO/STO heterostructure containing various atomic-scale thicknesses of the SRO and STO layers and monitored the structure using scanning transmission electron microscopy (STEM) and electron energy loss spectroscopy (EELS) (see Figures S5 and S6, and Methods). Figure 2(a) and 2(b) show the cross-sectional STEM images of the artificial SRO/STO heterostructure acquired in the high-angle annular dark field (HAADF) imaging mode, in which the SRO layers show a brighter band contrast owing to the dependence of intensity approximately scaled to the atomic number squared.[36] The low-magnification HAADF STEM image taken along the [110] direction of the STO substrate shows the well-controlled layer thickness of the SRO/STO heterostructures (Figure 2(a)). Figure 2(b) displays the high-magnification HAADF STEM result for the orange rectangular region in Figure 2(a), demonstrating well-ordered atomic columns (Figures S7 and S8).

The local structural anisotropy increases with decreasing *x* via lattice elongation. To capture the weakly scattered signals from the oxygen atoms, we examined the artificial heterostructure in the annular bright field (ABF) STEM imaging mode, which allows effective collection of the signal of a low-atomic-number element,[37] as presented in Figure 2(c). We focused on the 6 u.c. SRO layers (#1, upper panel) and 2 and 4 u.c. of SRO layers (#2, lower panel) sandwiched between identical 8 u.c. STO layers. The ABF STEM images show a projected heterostructure oriented in the [110] direction of the STO substrate. The black rectangles denote the SRO layers, and the red rectangle in the upper panel of Figure 2(c) shows that all the Sr (green), Ti or Ru (blue), and O (red) atoms within the perovskites were located at their respective positions without any crystallographic defects. Figure 2(d) shows lattice-elongation maps for the [110]-oriented perovskite u.c. along the in-plane (left panel) and out-of-plane (right panel) directions. The false color contrast of each pixel in those maps indicates the relative magnitude of the projected Sr–Sr atomic distance for the in-plane and out-of-plane directions, which were compared with the [110]-oriented STO cubic bulk-structure values (2.765 and 3.905 Å for the in-plane and out-of-plane directions, respectively). The lattice elongation did not change significantly along the in-plane direction, indicating an epitaxially strained state, which is consistent with the XRD-RSM analyses. In contrast, we observed finite and systematic lattice elongation along the out-of-plane direction, particularly for the atomically thin SRO layers. The out-of-plane lattice profiles in Figure 2(e) reveal the considerable lattice elongation in SRO layers within the heterostructure reaching up to 16 pm for the 2 u.c. SRO layers. The average values of the lattice elongation were 7.59, 8.62, and 9.87 pm for the 6, 4, and 2 u.c. SRO layers, respectively. The systematic suppression of out-of-plane lattice elongation as a function of the SRO-layer thickness experimentally demonstrated lattice deformations for the atomically thin

SRO layers owing to octahedral coupling within the SRO/STO heterostructures, despite the substrate-clamping effect. The dissimilar octahedral distortions of bulk orthorhombic SRO and cubic STO may result in additional out-of-plane lattice elongation without changing the in-plane strain state.[34]

The second-harmonic generation (SHG) results support the enhancement of structural anisotropy with decreasing $x$ in the SRO/STO superlattices, consistent with our atomic-scale structural analyses (Figures 3(a) and 3(b)). In principle, the SHG signal ($I_{SHG}$) is proportional to the square of the electric-dipole polarization, which is sensitive to crystalline symmetry. Especially, whereas thick SRO films have negligible $I_{SHG}$ because of their centrosymmetric nature, local structural distortion with inversion symmetry breaking in thin SRO films enhances $I_{SHG}$ significantly.[38-40] We measured the $I_{SHG}$ of SRO superlattices as a function of the in-plane azimuthal rotation angle ($\phi$) in a normal-incidence geometry with parallel (XX) polarization and a grazing-angle configuration with PP polarizations (see Methods), as schematically shown in the lower panels of Figure 3(a). While $I_{SHG}$ ($\phi$) patterns of SRO superlattice with different $x$ showed the same rotational symmetry, the $I_{SHG}$ amplitude significantly and systematically changed as $x$ varied. As summarized in Figure 3(b), the relative area of $I_{SHG}$ ($\phi$) patterns for the XX configuration significantly increases by more than three times as $x$ decreases from six to one. Interestingly, $I_{SHG}$ for PP configuration exhibits an opposite $x$-dependence. Considering that the in-plane and out-of-plane polarizations are sensitively probed in XX and PP configurations, respectively, the opposite $x$-dependence of $I_{SHG}$ for two different configurations implies the systematic variation of the structural anisotropy (tetragonality). Whereas the SHG result alone would be insufficient to quantitatively determine the lattice elongation, the consistent behavior observed from the STEM analyses ensures the dimensionality-driven anisotropy evolution. The octahedra shown in Figure 3(b) are schematically expected lattice elongation along the out-of-plane direction for the low-dimensional SRO layers.

Figures 3(c) and 3(d) show the X-ray absorption spectroscopy (XAS) results, which are used to observe variations in the anisotropic Ru-O hybridization, providing a possible mechanism for $x$-dependent MA modulation. The XAS spectrum of perovskite oxides at the O $K$-edge reflects the orbital hybridization between the unoccupied oxygen 2$p$ and transition-metal orbital states.[41] As shown in Figure 3(c), the O $K$-edge XAS spectra of the SRO/STO superlattice with 0° (beam normal, right panel) and 70° (left panel) incidence angles show a peak at ~528.4 eV, indicating the Ru 4$d$ ($t_{2g}$)-O 2$p$ orbital hybridization in the SRO layers. The strong peak at ~530.4 eV originates from Ti 3$d$-O 2$p$ orbital hybridization. For comparison, the spectra were normalized by the prominent Ti 3$d$ peak intensity. To enhance the visibility of the Ru 4$d$-related signals, we subtracted the Ti 3$d$ peaks assuming Lorentzian line shapes (gray peak) in Figure 3(c). The Ru 4$d$-O 2$p$ peak intensities for both angles decrease with decreasing SRO thickness, and for $x = 1$, the peak becomes almost invisible. This result supports the SRO-thickness-dependent ferromagnetic metal-to-insulator transition and consequent electron localization,[14] analogous to a previous study on SRO single films.[41] To assess the $x$-dependent anisotropy of Ru-O hybridization, we normalized the peak intensities of the X-ray linear dichroism (XLD) at 528.4 eV. Specifically, the difference between the horizontally and vertically polarized XAS components (E//c – E//ab) was divided by the number of SRO layers within the superlattice (see Methods for more details), and the results are summarized in Figure 3(d). The XAS intensities of the vertically and horizontally polarized components reflect the Ru 4$d$ ($t_{2g}$) orbital hybridization with apical O 2*p* orbitals and the sum of 1/2 equatorial and 1/2 apical O 2*p* orbitals, respectively. Therefore, a positive XLD signal indicates a larger orbital overlap of Ru compounds with equatorial O ions than with apical O ions. The weak XLD intensity of the *x* = 6 superlattice supported the relatively isotropic orbital hybridization of the thick SRO layers. Moreover, as *x* decreased to 3, the XLD intensity increased by more than nine times compared with the case of the *x* = 6 superlattice. This suggests enhanced anisotropy in the Ru-O orbital hybridization, that is, the ratio of the Ru hybridization strength of equatorial O ions to that of apical O ions significantly increased for atomically thin SRO layers (Figure 3(d)). As *x* further decreased, the dimensional crossover effect from 3D to 2D and electron localization suppressed the XLD signals and orbital hybridization.[14] Interestingly, the *x*-dependent XLD intensity followed the same trend as the magnetic $H_c$ at low temperatures (Figure 1(e)). This correspondence implies that anisotropic Ru-O orbital hybridization is closely related to the *x*-dependent MA in the SRO layers.[23, 30]

We demonstrated the dimensionality and tetragonality modulation of the ferromagnetic order and its magnetic anisotropy using atomically controlled correlated oxides. By taking advantage of the periodic supercell structure, we examined the dimensional crossover in correlated magnetic SRO at the atomic-scale monolayer limits. Precise atomic-scale control leads to a nonmonotonic engineering of $H_c$ obtained from anomalous Hall effects with an ~1,500% improvement for 3 u.c. SRO layers compared to SRO single film. However, the dimensional crossover of ferromagnetism alone does not describe the MA modulation. To understand the nonmonotonic variation near the monolayer limits, we designed synthetic SRO/STO heterostructures with atomic-scale precision and confirmed the intriguing local structural anisotropy near the two-dimensional limits using both STEM and SHG measurements. These results lead to a sizeable anisotropy modulation in Ru-O orbital hybridizations established by XAS, suggesting a possible origin for the significant enhancement of MA in atomically thin SRO layers. Our observations provide a systematic tunability of the correlated long-range spin order across the dimensional crossover. The approach can be implemented in various low-dimensional correlated materials to investigate the anisotropy and ground states of synthetic magnetic crystals.

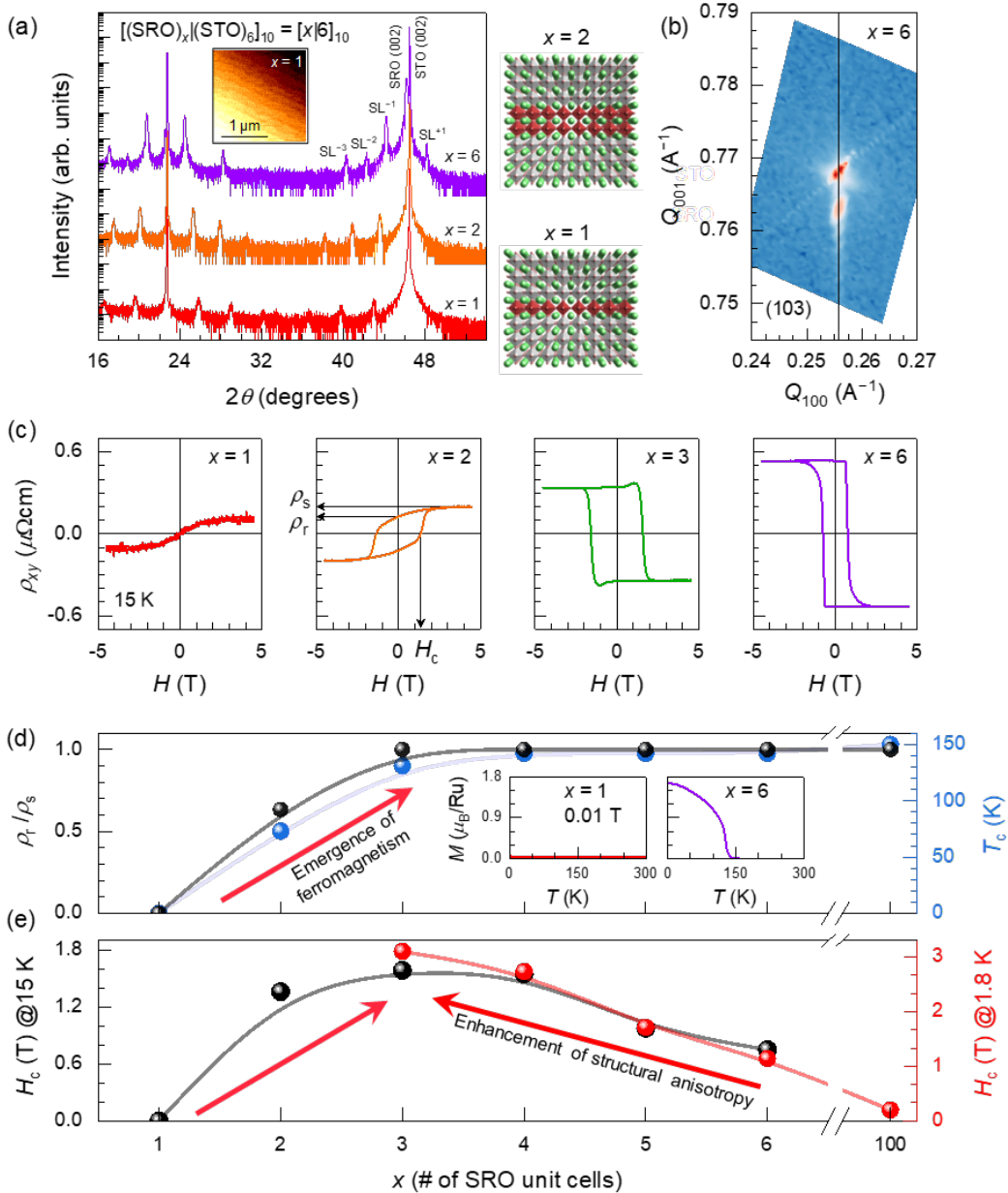

**Figure 1.** Dimensionality-controlled magnetic anisotropy in atomically designed magnetic superlattices. (a) XRD $\theta$-$2\theta$ measurements of SRO/STO superlattice for 6 u.c. STO layer with different $x$ values, i.e., $[x|6]_{10}$. SL$^{\pm n}$ denotes the superlattice peak positions. The inset represents the atomic force microscopy image of the $x = 1$ superlattice. The schematic representation on the right shows the supercell structure of the superlattice for a few atomic SRO layers ($x = 1$ and 2). (b) XRD RSM results around the (103) plane of STO for $x = 6$ superlattice, as an example. (c)

Out-of-plane $H$-field-dependent $\rho_{xy}$ of superlattices with different $x$ measured at 15 K. The $\rho_r$ and $\rho_s$ values were obtained at $H$-fields of 0 and 4 T. (d) As $x$ decreases to a limit of a few atomic layers, the remanence ratio ($\rho_r/\rho_s$, black) and $T_c$ (blue) consistently decrease, representing a suppression of the ferromagnetic behavior in SRO layers at a two-dimensional limit. The insets show the temperature-dependent out-of-plane magnetization for $x = 1$ and 5 superlattices with an out-of-plane magnetic field of 100 Oe. (e) As $x$ decreases, the $H_c$ values at 15 K (black) and 1.8 K (red) increase up to $x = 3$ owing to the enhancement of structural anisotropy. However, when $x < 3$, $H_c$ decreases with decreasing $x$ due to the dimensional-crossover effect. The $H_c$ value of $x = 100$ at 1.8 K was adopted for SRO single film from a previous study.[34, 35]

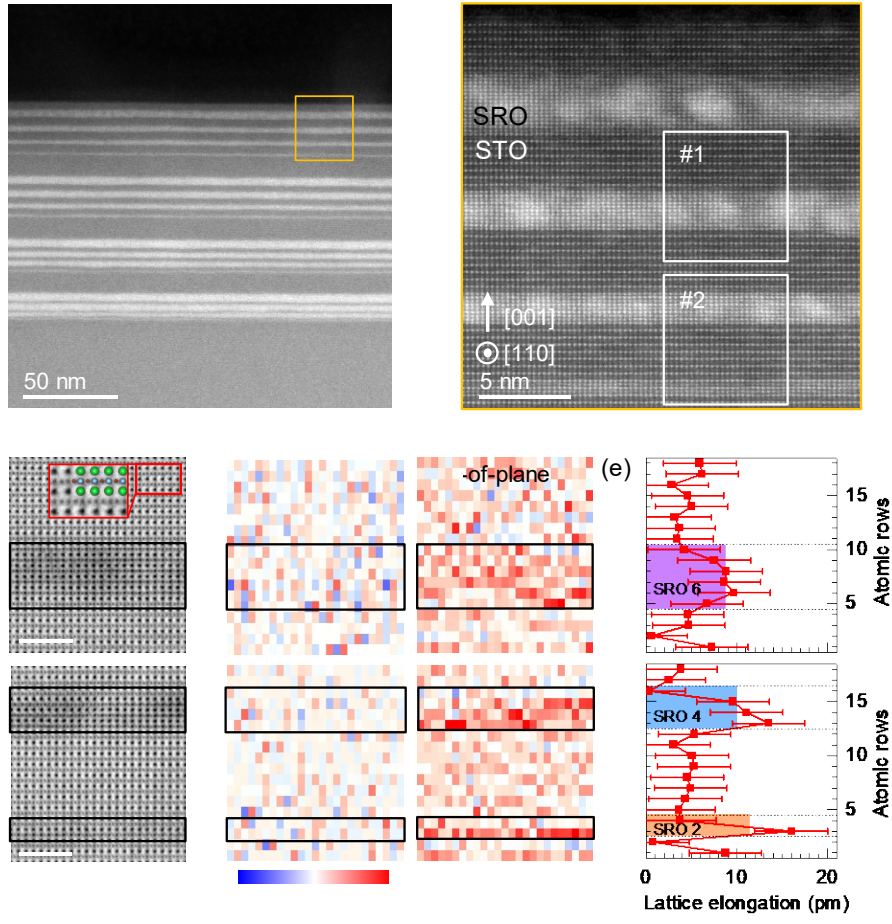

**Figure 2.** Enhancement of local structural anisotropy across the dimensional crossover. Cross-sectional HAADF STEM images of the SRO/STO heterostructure comprising different perovskite u.c. of the constituent materials in (a) low and (b) high magnifications observed along the [110] direction of the substrate. The number of atomic layers was resolved to confirm the designed structure, and the interfaces were proven to be atomically sharp. (c) ABF STEM images of the SRO/STO heterostructure in high magnification. (Region #1 and #2 are marked by white rectangles in (b).) The black rectangles indicate the SRO layers. (d) Lattice-elongation maps for the perovskite u.c. for the in-plane (left) and out-of-plane (right) directions. (e) The out-of-plane lattice elongation, averaging 25 data points in the same row, was obtained along the growth directions.

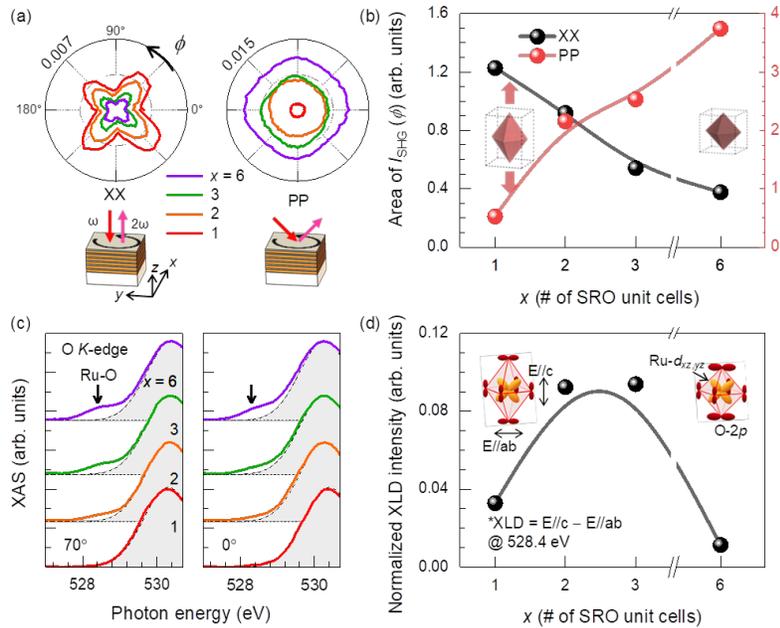

**Figure 3.** Structural-anisotropy enhancement and resultant Ru-O orbital hybridization of atomically thin SRO layers within superlattices. (a) $\phi$-dependence of SHG signal measured in front-reflection geometry with XX (left panel) and PP (right panel) polarization. The schematic shows the normal (grazing) incident configuration for XX (PP) configuration employed for SHG measurements. (b) As $x$ decreases, the area of the SHG rotational-anisotropy pattern of the XX channel (black symbols) increases, but that of the PP channel (red symbols) decreases. These represent an improvement in structural anisotropy for the atomically thin SRO layers. The inset shows schematic illustrations of the expected enhancement of structural anisotropy, which is an exaggerated representation compared to the observed results. (c) O $K$-edge XAS spectra and (d) normalized XLD (E//c – E//ab) intensities at 528.4 eV of SRO/STO superlattices with different $x$ values. The left (right) panel of (c) is the spectra with a 70° (0°) incident angle. The gray Lorentzian peaks account for the contribution of Ti 3$d$-O 2$p$ orbital hybridizations. The inset shows a schematic of anisotropic Ru 4$d$-O 2$p$ orbital hybridizations for the atomically thin SRO layers.

## ASSOCIATED CONTENT

The Supporting Information is available free of charge at

Details of Methods for atomic-scale epitaxial growth, lattice structure characterization, Electrical and magnetization measurements, second harmonic generation, X-ray absorption spectroscopy, and Figures S1–S18 (PDF).

## AUTHOR INFORMATION

**Corresponding Author**

**Woo Seok Choi** − *Department of Physics, Sungkyunkwan University, Suwon 16419, Korea*; https://orcid.org/0000-0002-2872-6191; E-mail: choiws@skku.edu

**Deok-Yong Cho** − *IPIT & Department of Physics, Jeonbuk National University, Jeonju 54896, Korea*; E-mail: zax@jbnu.ac.kr

**Young-Min Kim** − *Department of Energy Sciences, Sungkyunkwan University, Suwon 16419, Korea*; E-mail: youngmk@skku.edu

**Suyoun Lee** − *Center for Neuromorphic Engineering, Korea Institute of Science and Technology, Seoul 02792, Korea*; E-mail: slee_eels@kist.re.kr

**Author Contributions**

S.G.J. and S.W.C. contributed equally to this work.

**Notes**

The authors declare no competing financial interest.

**Acknowledgments**

This work was supported by the Basic Science Research Program through the National Research Foundation of Korea (NRF-2021R1A2C2011340, 2022M3H4A1A04074153, RS-2023-00220471, RS-2023-00281671, 2022R1C1C2006723, 2023R1A2C2002403, and 2021M3F3A2A03017782). S.W.C. and S.L. were financially supported by the Korea Institute of Science and Technology (KIST) through 2E32960.